\documentclass{vow2008}

\title{Spectral Analysis in the Virtual Observatory}
\author{Thomas Rauch}
\affil{Institute for Astronomy and Astrophysics,
       Kepler Center for Astro and Particle Physics,
       Eberhard Karls University, 
       Sand 1,
       72076 T\"ubingen, 
       Germany}
\author{Iliya Nickelt}
\affil{Astrophysical Institute Potsdam,
       Department of E-Science,
       An der Sternwarte 16, 
       14482 Potsdam, 
       Germany}

\usepackage{color}
\usepackage{graphicx}

\begin{document}


\maketitle

\begin{abstract}
In a collaboration of the German Astrophysical Virtual 
Observatory (\emph{GAVO}) and \emph{AstroGrid-D}, the German Astronomy Community 
Grid (GACG), we provide a VO service for the access and the 
calculation of stellar synthetic energy distributions (SEDs) based 
on static as well as expanding non-LTE model atmospheres.

At three levels, a VO user may directly compare observed and 
theoretical SEDs: The easiest and fastest way is to use 
pre-calculated SEDs from the GAVO database. For individual objects, 
grids of model atmospheres and SEDs can be calculated on the compute 
resources of \mbox{\emph{AstroGrid-D}} within reasonable wallclock time. 
Experienced VO users may even create own atomic-data files for a 
more detailed analyses.
\end{abstract}

\section{Introduction}
\label{sect:introduction}

Spectral analysis of hot, compact stars by means of Non-LTE (local
thermodynamic equilibrium) model-atmosphere techniques has for a 
long time been regarded as a domain of specialists. In contrast to
the assumption of LTE, where occupation numbers of atomic levels
are determined by Saha equation and Boltzmann statistics, radiative
and collisional transitions have to be considered in detail 
\citep[see, e.g.,][]{w1986}. This
makes the calculation of an elaborated model and its spectral energy 
distribution (SED) time-consuming, although it can be performed
on presently available PCs. Faster and easier is the use of blackbody
SEDs, tempting, e.g., old-fashioned users of photoionization codes 
to use such for the (bad) representation of the exciting source of 
an ionized gaseous nebula -- spoiling the advantages of the best 
of these codes.

Within the last four decades, Non-LTE model-atmospheres arrived at a 
high level of sophistication: fully line-blanketed models consider 
opacities of all elements from H to Ni \citep{r2003,rea2007}. 
Within the GAVO\footnote{http://www.g-vo.org}, 
we provide access to SEDs of such models (Sect.~\ref{sect:theossa}). 

In general, spectral analysis requires grids of SEDs over a wide parameter
range (effective temperature $T_\mathrm{eff}$, surface gravity $g$, element abundances). 
Since their calculation may exceed the available capabilities, model
atmosphere calculations are performed on compute resources of
\emph{AstroGrid-D}.

\begin{figure*}[ht]
\begin{picture}(17.0,10.0)
\thicklines

\put( 3.5, 8.0){\makebox(6.0,1.0)[c]{{\color{red}\emph{{\large TheoSSA} request:} $T_\mathrm{eff}$, $\log g$, $X_\mathrm{i}$}}}
\put( 4.0, 8.5){\oval( 1.0, 1.0)[l]}
\put( 9.0, 8.5){\oval( 1.0, 1.0)[r]}
\multiput( 4.0, 8.0)(0.0,1.0){2}{\line(1,0){5.0}}

\put( 6.5, 8.0){\vector(0,-1){0.5}}

\put( 4.5, 6.25){\makebox(4.0,1.0)[c]{{\begin{minipage}{50mm}
                                      \begin{center}
                                      check {\color{blue}GAVO database}:\\
                                      requested parameters matched\\
                                      within tolerance limits?
                                      \end{center}
                                      \end{minipage}
                                      }}}
\put( 4.5, 6.75){\oval( 1.0, 1.5)[l]}
\put( 8.5, 6.75){\oval( 1.0, 1.5)[r]}
\multiput( 4.5, 6.0)(0.0,1.5){2}{\line(1,0){4.0}}

\put( 6.5, 6.0){\line(0,-1){0.25}}
\put( 3.75, 5.75){\line(1,0){5.5}}
\multiput( 3.75, 5.75)(5.5,0.0){2}{\vector(0,-1){0.25}}

\multiput( 2.0, 5.0)(5.0,0.0){2}{\oval( 1.0, 1.0)[l]}
\multiput( 5.5, 5.0)(6.0,0.0){2}{\oval( 1.0, 1.0)[r]}
\multiput( 2.0, 4.5)(0.0,1.0){2}{\line(1,0){3.5}}
\multiput( 7.0, 4.5)(0.0,1.0){2}{\line(1,0){4.5}}
\put( 1.75, 4.5){\makebox(4.0,1.0)[c]{{yes: offer existing SED}}}
\put( 7.25, 4.5){\makebox(4.0,1.0)[c]{{\begin{minipage}{50mm}
                                      \begin{center}
                                      no: calculate new model and SEDs\\
                                      and ingest into {\color{blue}GAVO database}
                                      \end{center}
                                      \end{minipage}
                                      }}}

\put( 3.75, 4.5){\line(0,-1){0.25}}
\put( 1.75, 4.25){\line(1,0){4.0}}
\put( 1.75, 4.25){\vector(0,-1){1.25}}
\put( 5.75, 4.25){\vector(0,-1){0.25}}

\put( 0.5, 2.50){\oval( 1.0, 1.0)[l]}
\put( 3.0, 2.50){\oval( 1.0, 1.0)[r]}
\multiput( 0.5, 2.00)(0.0,1.0){2}{\line(1,0){2.5}}
\put( 0.25, 2.00){\makebox(3.0,1.0)[c]{{accept, retrieve SED}}}
\put(13.50, 2.50){\vector(-1,0){10.00}}

\put( 4.5, 3.50){\oval( 1.0, 1.0)[l]}
\put( 7.75, 3.50){\oval( 1.0, 1.0)[r]}
\multiput( 4.5, 3.00)(0.0,1.0){2}{\line(1,0){3.25}}
\put( 3.85, 3.0){\makebox(4.5,1.0)[c]{{request exact parameters}}}
\put( 8.25, 3.5){\vector(1,1){1.0}}

\put(14.0, 7.75){\oval( 1.0, 3.0)[l]}
\put(17.0, 7.75){\oval( 1.0, 3.0)[r]}
\multiput(14.0, 6.25)(0.0,3.0){2}{\line(1,0){3.0}}
\put(13.5, 6.25){\makebox(4.0,3.0)[c]{{\begin{minipage}{40mm}
                                      \begin{center}
                                      {\color{blue}GAVO database @ ARI}\vspace{5mm}\\
                                      {\color{red}$T_\mathrm{eff}$, $\log g$, $X_\mathrm{i}$}\vspace{5mm}\\
                                      metadata
                                      \end{center}
                                      \end{minipage}
                                      }}}

\put( 9.0, 6.75){\vector(4,1){4.5}}
\put(13.5, 7.8775){\vector(-4,-1){4.5}}
\put(12.0, 5.00){\vector(1,1){1.5}}

\put(14.0, 3.00){\oval( 1.0, 3.0)[l]}
\put(17.0, 3.00){\oval( 1.0, 3.0)[r]}
\multiput(14.0, 1.50)(0.0,3.0){2}{\line(1,0){3.0}}
\put(13.5, 1.5){\makebox(4.0,3.0)[c]{{\begin{minipage}{40mm}
                                      \begin{center}
                                      {\color{blue}GAVO database @ IAAT}\vspace{2mm}\\
                                      {\color{red}$T_\mathrm{eff}$, $\log g$, $X_\mathrm{i}$}\vspace{2mm}\\
                                      models\\
                                      atomic data\\
                                      frequency grids\\
                                      SEDs
                                      \end{center}
                                      \end{minipage}
                                      }}}

\put(12.0, 5.00){\vector(2,-1){1.5}}

\put(15.5, 4.50){\vector(0,1){1.75}}
\put(15.5, 6.25){\vector(0,-1){1.75}}

\end{picture}
\caption{Data flow within \emph{TheoSSA}. ARI: Astronomisches Recheninstitut (Heidelberg, Germany), IAAT:
Institut f\"ur Astronomie und Astrophysik T\"ubingen (Germany), $X_\mathrm{i}$: 
mass fraction of species $\mathrm{i} \in \left[\mathrm{H, He, C, N, O}\right]$.}
\label{fig:dataflow}
\end{figure*}
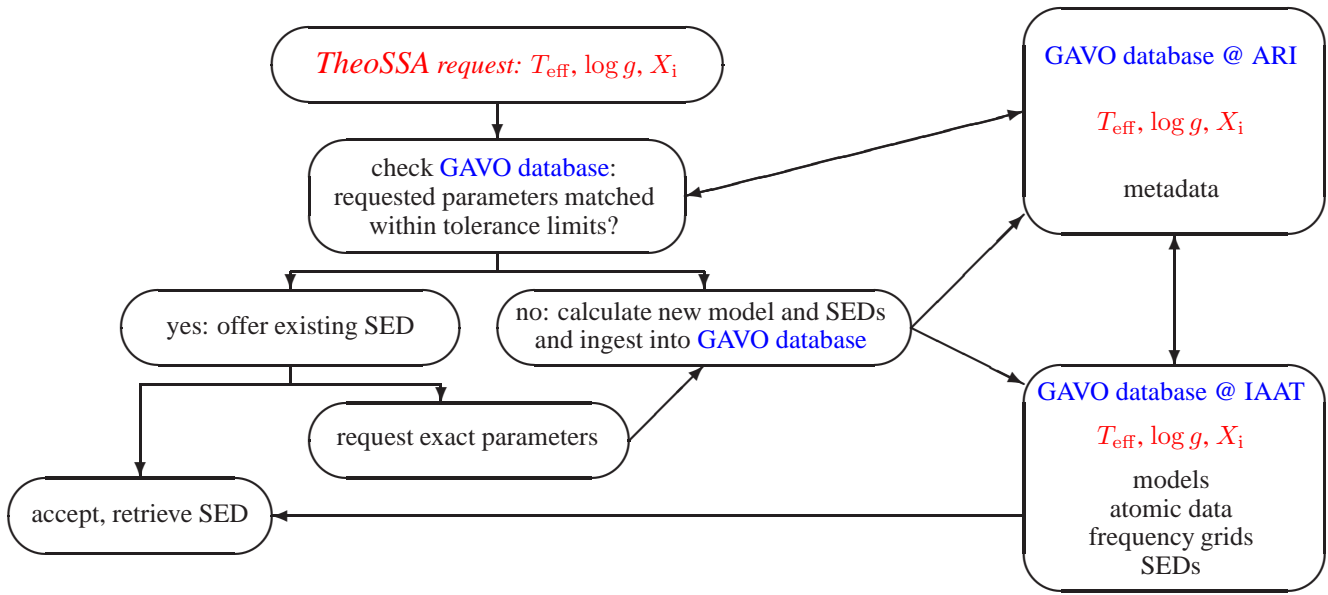

\section{TheoSSA -- Model SED on Demand}
\label{sect:theossa}

We have created the VO service \emph{TheoSSA} 
(Theoretical Simple Spectra Access\footnote{http://vo.ari.uni-heidelberg.de/ssatr-0.01/TrSpectra.jsp}. 
At three levels
it provides access to SEDs that are in a pilot phase calculated by the T\"ubingen Model Atmosphere Package
\emph{TMAP}\footnote{http://astro.uni-tuebingen.de/\raisebox{0.3em}{\tiny$\sim$}rauch/TMAP/TMAP.html}
\citep{wea2003,rd2003} for hot, compact stars only:\\

{\bf Fast and Easy}\\ An unexperienced VO user (no detailed knowledge about the model-atmosphere code necessary) may
                      use pre-calculated SEDs, interpolate in between.\vspace{2mm}\\
{\bf Individual}   \\ An interested VO user can calculate own SEDs for the analysis of special objects.\vspace{2mm}\\
{\bf Experienced}  \\ The VO user can define own model atoms and upload them for the comparison with other codes, etc.\vspace{2mm}

Figure~\ref{fig:dataflow} shows the scheme of the dataflow within \emph{TheoSSA}.
If individual parameters are requested by a VO user, the GAVO database is checked for suitable SEDs (parameters match within
error limits). If there is no match, the VO user is guided to
\emph{TMAW}\footnote{http://astro.uni-tuebingen.de/\raisebox{0.3em}{\tiny$\sim$}TMAW/TMAW.shtml}.
With this WWW interface, the VO user may calculate an individual model 
atmosphere, requesting effective temperature, surface gravity,
and mass fractions of H, He, C, N, and O (more species will be included 
in the future). For this calculation, standard model atoms are used which 
are provided within the T\"ubingen Model-Atom Database 
\emph{TMAD}\footnote{http://astro.uni-tuebingen.de/\raisebox{0.3em}{\tiny$\sim$}rauch/TMAD/TMAD.html}.
Since the VO user can do this without detailed knowledge of the programme 
code working in the background, the access to individually calculated SEDs 
is as simple as the use of pre-calculated SEDs -- however,
the calculation needs some time (depending on the number of species 
considered, the wall-clock time is ranging from hours to a few days). 
Standard SEDs of all calculated model atmospheres are automatically ingested 
into the GAVO data base and, thus, it is growing in time.

If a detailed spectral analysis is performed, an experienced VO 
user may create an own atomic data file tailored for a specific purpose 
considering all necessary species and calculate own model atmospheres and SEDs.

\section{TheoSSA@Grid}
\label{theossaatgrid}

\emph{AstroGrid-D}\footnote{http://www.gac-grid.de/}
is a German research and development project to build an astronomical research
infrastructure. Using the Globus Toolkit (GT4) middleware, \emph{AstroGrid-D} embeds existing 
computational facilities, dedicated resources and specialized hardware, such as robotic telescopes. 
Almost a hundred users from the German astronomical community use its compute and storage facilities.
The project also resulted in several scientific applications and novel Grid services, such as the
information service \emph{Stellaris}\footnote{http://stellaris.zib.de/}.

The \emph{TMAP} model calculations are carried out on the resources of \emph{AstroGrid-D} to ensure a reliable service, 
even for a high number of requests. For job submission, a script package is used that 
allows to submit jobs to computational resources automatically. This script package was originally 
developed for a different \emph{AstroGrid-D} use 
case\footnote{http://www.gac-grid.org/project-products/Applications/}, but can be easily
adopted to other implementations of so-called \emph{atomic grid jobs}. The term refers to a specific
type of grid application, where all input data is transferred together with the software without
need for inter-process communication. The grid is thus used as a task farming mechanism. 
But even if this is only a comparatively simple 
application for the use of the flexible GT4 middleware, we found task farming to be an 
important requirement of many use cases, such as \emph{TMAP}.

The transfer of the \emph{TMAP} model software and the input data, as well as the transmission of the 
results is part of the standard job submission process in Globus Toolkit. We make use of the 
GT4 web services and control the process by a template written in a JSDL (Job Submission 
Description Language)\footnote{http://www.gridforum.org/documents/GFD.56.pdf}. Whenever a user requests 
new model data to be calculated, the template is 
adopted to the specific case and a target machine is selected from a given list. Upon arrival,
the results are passed on to the \emph{TheoSSA} service to notify the user.

\section{Conclusions and Future Plans}
\label{summary}

\emph{TheoSSA} is already fully functional and provides easy access to SEDs of hot stars -- use them!
E.g\@. the use of blackbody SEDs to represent stars in photoionization models belongs to the
last millennium and is not adequate anymore. 

The database of complete model atoms for other species (Ne, Na, Mg, Si, ...) 
up to the iron group will be extended. \emph{TMAW} can already consider opacities of all elements
from hydrogen to germanium.

The addition of other model codes for other types of stars will in future further improve the 
benefit for VO users. This is an invitation and challenge for other stellar-atmosphere working groups
to contribute.

Precise spectral analysis requires extended grids of elaborated model atmospheres.
In the framework of GRID computing \citep{f2005, f2006}, 
the calculation of model-atmosphere grids and flux tables, e.g\@. via \emph{TheoSSA}, is an excellent 
application to efficiently calculate synthetic spectra on reasonable time scales.

One approach is already in preparation, using the newly developed \emph{HotBlast} 
Non-LTE code for spherically expanding stellar atmospheres (Koesterke priv.~comm.). 
\emph{HotBlast} uses as an input the atmospheric structure of the static \emph{TMAP} model atmospheres 
to simulate the atmosphere below the wind region.

\section*{Acknowledgments}

We thank the \emph{GAVO} and \emph{AstroGrid-D} teams for support.
The \emph{German Astrophysical Virtual Observatory} (GAVO) project is sponsored by the 
German Federal Ministry of Education and Research (BMBF) under grants 05\,AC6VTB and 05\,AC6VHA.
\emph{AstroGrid-D} ist sponsored by the German Federal Ministry of Education and Research within 
the \emph{D-Grid initiative} under contracts 01\,AK804[A-G].

\end{document}